\documentclass[letterpaper]{article} 
\usepackage{aaai2026}  
\usepackage{times}  
\usepackage{helvet}  
\usepackage{courier}  
\usepackage[hyphens]{url}  
\usepackage{graphicx} 
\usepackage{natbib}  
\usepackage{caption} 
\usepackage{algorithm}
\usepackage{algorithmic}

\usepackage{newfloat}
\usepackage{listings}
\usepackage{amsmath}
\usepackage{booktabs}
\DeclareCaptionStyle{ruled}{labelfont=normalfont,labelsep=colon,strut=off} 
\floatstyle{ruled}
\newfloat{listing}{tb}{lst}{}
\floatname{listing}{Listing}
\title{Driving Engagement in Daily Fantasy Sports with a Scalable and Urgency-Aware Ranking Engine}
\author {
    Unmesh Padalkar
}
\affiliations{
    Dream11\\
    unmesh.padalkar@dream11.com
}
\begin{document}

\maketitle

\begin{abstract}
In daily fantasy sports (DFS), match participation is highly time-sensitive. 
Users must act within a narrow window before a game begins, making match recommendation a time-critical task to prevent missed engagement and revenue loss. Existing recommender systems, typically designed for static item catalogs, are ill-equipped to handle the hard temporal deadlines inherent in these live events. To address this, we designed and deployed a recommendation engine using the Deep Interest Network (DIN) architecture. 
We adapt the DIN architecture by injecting temporality at two levels: first, through real-time urgency features for each candidate match (e.g., time-to-round-lock), and second, via temporal positional encodings that represent the time-gap between each historical interaction and the current recommendation request, allowing the model to dynamically weigh the recency of past actions. 
This approach, combined with a listwise neuralNDCG loss function, produces highly relevant and urgency-aware rankings. To support this at industrial scale, we developed a multi-node, multi-GPU training architecture on Ray and PyTorch. Our system, validated on a massive industrial dataset with over 650k users and over 100B interactions, achieves a +9\% lift in nDCG@1 over a heavily optimized LightGBM baseline with handcrafted features. The strong offline performance of this model establishes its viability as a core component for our planned on-device (edge) recommendation system, where online A/B testing will be conducted.
\end{abstract}

\begin{links}
\end{links}

\section{Introduction}\label{sec:introduction}
\begin{figure}[t]
    \centering
    \includegraphics[width=0.9\columnwidth]{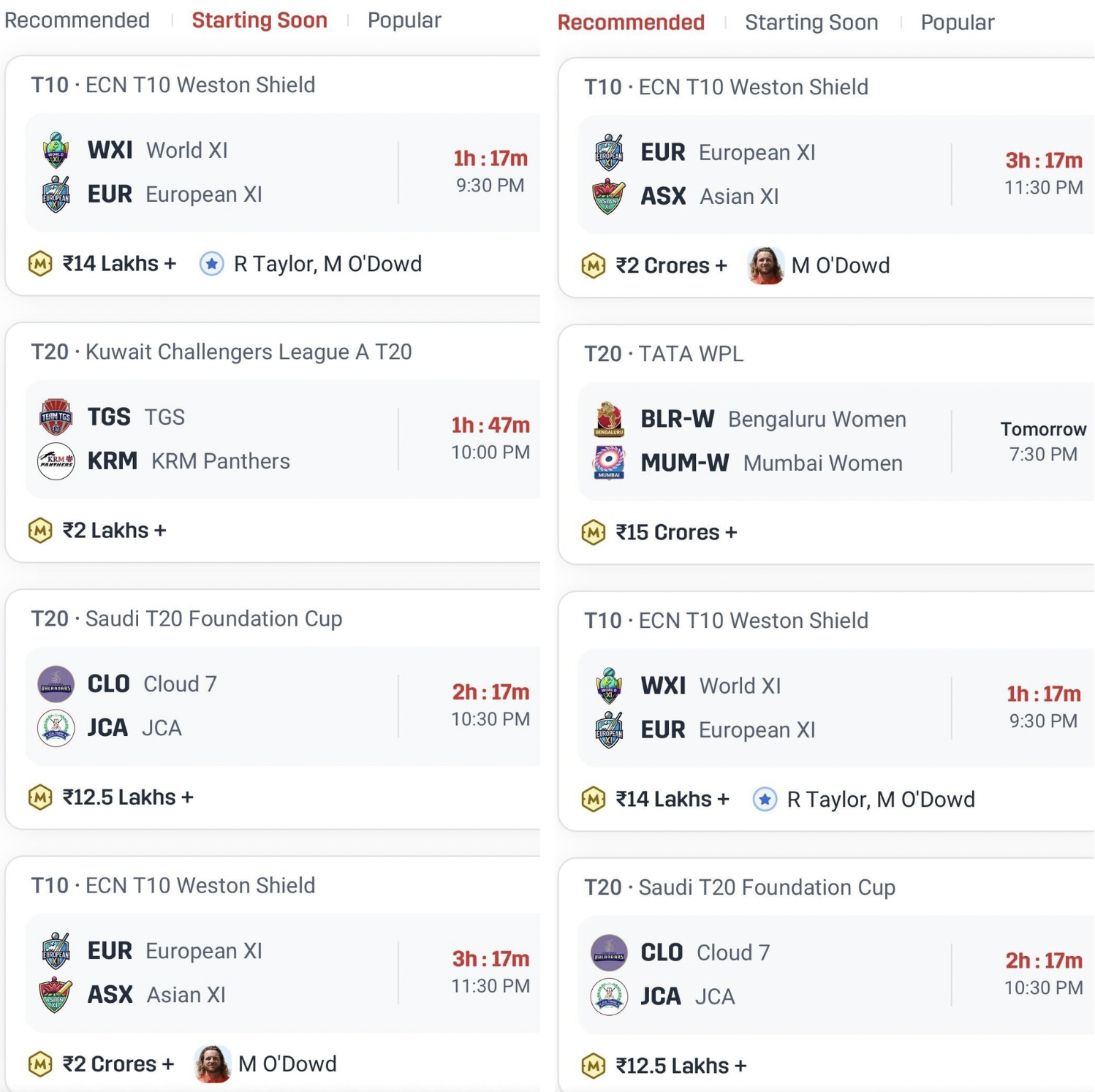}
    \caption{Comparison of match rankings: Starting Soon section ranks matches according to match start time, while Recommended section ranks matches based on machine learning model outputs.}
    \label{fig:mcp_ranking_homepage}
\end{figure}
The global fantasy sports market, valued at over USD 30 billion in 2023 and projected to reach nearly USD 85 billion by 2032, represents a massive and rapidly growing digital entertainment sector \cite{fantasy_sports_market}. 
Our work is situated within the Daily Fantasy Sports (DFS) format, a rapidly expanding segment characterized by short-term games, typically spanning a few hours to a few days. 
In DFS, users act as virtual managers, creating teams of real-world players to compete based on those players' statistical performance in a live match. 
At the heart of major fantasy sports platforms, the homepage list of upcoming matches (Figure \ref{fig:mcp_ranking_homepage}) serves as the primary gateway to user engagement. 
While this can be framed as a Learning-to-Rank (LTR) problem, it presents a unique challenge in balancing user relevance with what we term temporal urgency. 
For instance, a highly anticipated match starting in a week is relevant, but a less popular match that begins in one hour has a higher urgency. 
This urgency is critical from a business perspective, as the platform may bear financial losses for any contests that remain unfilled when a match starts. 
This creates a constant trade-off: prioritizing only user relevance risks revenue loss from expired matches, while prioritizing only urgency can flood users with irrelevant content, harming long-term engagement and user satisfaction
This challenge is amplified by a diverse user base, which includes both recreational fans and others who behave as utility-maximizing agents, strategically selecting and sequencing matches to optimize expected returns on investment (ROI). 
Crucially, this entire process is governed by a hard deadline. A key trigger for user activity is the release of official player lineups, often just minutes before a match begins, after which team submissions are locked. 
This problem is further complicated by the heterogeneity of match formats, ranging from five-day Test matches to short-form T10 games, each with unique rules and user followings.

Standard recommendation models often fail to navigate these complex challenges effectively. Our own internal journey confirms these difficulties. 
Initial attempts to rank matches directly by historical conversion metrics created severe feedback loops that starved new content of exposure. 
Subsequent models suffered from a fundamental inference-training discrepancy – sample selection bias \cite{sample_selection_bias} and data sparsity \cite{data_sparsity} as they were trained on sparse, post-click user actions (e.g., contest joins) but were required to rank items at the pre-click impression stage. These limitations, along with the non-scalability of temporary solutions like control-group-based ranking, underscored the need for a more sophisticated, sequence-aware architecture.To solve this, we designed and built an urgency-aware recommendation engine centered on adapting the Deep Interest Network (DIN) \cite{din} architecture. This choice is directly motivated by the user's journey on the platform, which is inherently sequential: users browse matches, draft a team under a set of constraints (e.g., a salary cap), and finally, join a paid contest. DIN is naturally suited to model such behavioral sequences.

We demonstrate the effectiveness of our system through extensive offline experiments on a massive industrial dataset from our platform. Our model achieved a +9\% increase in nDCG@1 over a heavily optimized LightGBM \cite{lightgbm_paper} baseline, establishing its strong performance and readiness for online deployment. This work serves as the foundational modeling component for our next-generation on-device (edge) recommendation system, which is currently under development. The model is scheduled for live A/B testing within this new edge framework upon its completion.

Our contributions are as follows:
\begin{itemize}
\item We identify and formalize the critical trade-off between item urgency and user relevance in the multi-billion dollar Daily Fantasy Sports domain.
\item We propose an effective adaptation of the Deep Interest Network that models temporality in two key ways: (i) by incorporating real-time urgency features (e.g., time-to-round-lock) into the target item representation, and (ii) by adding temporal positional encodings to the user's historical interaction sequence.
\item We demonstrate the successful adaptation of the DIN architecture from its original pointwise (CTR prediction) formulation to a listwise ranking framework, successfully integrating it with the neuralNDCG \cite{neuralndcg} loss function for direct slate optimization.
\item We present a scalable multi-node, multi-GPU engineering framework on Ray \cite{ray_paper} and PyTorch \cite{pytorch_paper} for training on billions of industrial-scale interactions.
\item We validate our approach on a massive industrial dataset, showing significant improvements in offline ranking metrics over a strong LightGBM baseline and establishing its readiness for future online validation in a next-generation on-device (edge) environment.
\end{itemize}

\section{Related Work}\label{sec:related_work}
Our work is situated at the intersection of three core research areas: time- and urgency-aware recommendation, listwise learning-to-rank methodologies, and deep sequential user modeling.
Modeling temporal dynamics in recommender systems has a rich history.   
Classic approaches model the decay of user interest over time, emphasizing recent interactions \cite{koren_temporal_dynamics} or use time-based contextual features (e.g., timestamp embeddings) to capture seasonality and evolving preferences, a technique effectively demonstrated in large-scale industrial systems like YouTube's recommender \cite{youtube_recommendations}. 
More recent neural architectures, such as TiSASRec \cite{time_interval_aware_self_attention} and hidden semi‑Markov models \cite{hsmm}, explicitly encode periodicity and duration of user interest in sequential recommendations.
A sub-field known as urgency-aware recommendation explicitly considers items with limited availability windows—as in news recommendation, where relevance decays rapidly \cite{personalized_news_recommendation}, or coupon systems, where offers expire after a brief period \cite{coupon_rec}. 
In contrast, the DFS setting involves hard external deadlines (round lock time) and is further complicated by real-time events such as last‑minute lineup announcements—dynamics not fully addressed in prior work.

To solve these time-sensitive ranking challenges, the problem is naturally framed within the  Learning‑to‑Rank (LTR) problem with pointwise, pairwise, and listwise formulations \cite{ltr_survey}. Pointwise methods score each item independently (e.g., regression or classification). Pairwise methods, such as RankNet and LambdaRank/LambdaMART \cite{ranknet_lambdarank_lambdamart}, optimize for the correct ordering of pairs of items. Listwise methods, including ListNet \cite{listnet} and ListMLE \cite{listmle}, optimize a metric over the entire ranked list directly, aligning more closely with business objectives and the end-user experience.
Classic approaches like LambdaMART have been widely adopted in industrial settings via tree-based models (e.g., LightGBM, XGBoost) and are effective for ranking tasks with strong offline–online correlation. Recent deep learning works address the mismatch between training objectives and evaluation metrics by proposing differentiable relaxations of ranking metrics. neuralNDCG \cite{neuralndcg} offers a smooth, differentiable surrogate for the nDCG metric using a relaxed sorting operation, enabling end‑to‑end optimization of slate quality.

Within this LTR framework, effectively modeling the user's history is critical. Sequence-based models like the RNN-based GRU4Rec \cite{gru4rec} and the self-attention-based SASRec \cite{sasrec} are highly effective for next-item prediction, where the goal is to forecast a user's next interaction given their history. While these models can be extended to support candidate ranking and side features, such adaptations often require significant architectural changes.
In contrast, the Deep Interest Network (DIN) \cite{din} was explicitly designed for target-aware ranking in feature-rich environments. Its core innovation is a target-attention mechanism that computes a user interest vector conditioned on each candidate item independently, enabling dynamic, candidate-sensitive scoring. This distinction is particularly salient in the DFS setting, where a user’s interest in different sports (e.g., cricket vs. football) may activate different portions of their historical behavior. While SASRec would generate a uniform user embedding for all candidates, DIN recalibrates attention per candidate for more personalized, context-dependent scoring. This native support for rich features and target-centric attention makes DIN a more suitable and extensible foundation for our urgency-aware ranking task.

\section{Problem Formulation}\label{sec:problem_formulation}
The core task addressed in this paper is the real-time ranking of a slate of available matches for a given user to maximize engagement, defined as a click on a match. We approach this as a Learning-to-Rank (LTR) problem. 
For each observed match click (the positive sample), we construct a training instance by pairing it with a set of unchosen matches from the same context (negative samples). 
A fundamental challenge in this domain is that every match is a unique, ephemeral event with an ID that never repeats, creating an extreme item cold-start problem. 
Therefore, our model cannot rely on collaborative filtering from item IDs; it must be a content- and context-aware model that generalizes based on a rich set of item attributes (e.g., sport, teams involved, prize pool) and real-time signals (e.g.,  time-to-round-lock, lineup status).
The objective is to train this model to optimize a listwise ranking metric, such as nDCG, over this candidate set. 
The following subsections provide the formal definitions for this problem.

\subsection{The Ranking Task and Notation}
Let $\mathcal{U}$ be the set of all users and $\mathcal{I}$ be the set of all items (matches). 
At a specific request time $t_c$, a user $u \in \mathcal{U}$ is presented with a slate of available matches.
The set of candidate items for this ranking instance is denoted as $\mathcal{C}(u,t_c) = \{i_1, i_2, \ldots, i_m\}$.
This candidate set comprises all matches scheduled to start within a 24-hour window. 
Length of this window is a key production heuristic designed to solve a multi-objective trade-off: it focuses user attention on the most imminent matches, avoids locking a user's entry fee in a contest for a distant match until it concludes (potentially days later), thus preserving user liquidity for other matches, and maintains a computationally tractable candidate set for real-time ranking.
Within this context, the user interacts with a single target match $i^{+} \in \mathcal{C}(u,t_c)$, which serves as the positive label.
All other matches in the candidate set, $I^{-} = \mathcal{C}(u,t_c) \setminus \{i^{+}\}$, are treated as negative samples.

\subsection{User Interaction History}
A user's preferences are captured by their historical sequence of interactions. 
For each user $u$, we define their history as a time-ordered sequence of actions
$S_u = [a_1, a_2, \ldots, a_k]$, where each action $a_j$ occurred at a timestamp $t_j$.
Each action $a_j$ in the user's history sequence is a tuple $(i_j, t_j, \mathbf{x}_j)$, representing an interaction with match $i_j \in \mathcal{I}$ at timestamp $t_j$.
The associated feature vector $\mathbf{x}_j$ is multifaceted, designed to capture the full context of that historical event. 
It includes not only features representing the interaction type itself, but also the state of the match at that point in time, including its urgency (e.g., its time-
to-round-lock and lineup status relative to $t_j$) and contextual features (e.g., its prize pool).
Furthermore, the feature representation is enriched based on the type of interaction, which can be one of the following: 
\begin{itemize}
    \item \textbf{Match Click}: A baseline interaction indicating user interest, represented primarily by its type.
    \item \textbf{Team Save}: A stronger signal of a user's intent to participate in a match.
    \item \textbf{Contest Join}: The strongest signal of user commitment. This high-intent action is further enriched with specific features for the number of contests joined in that transaction and the total entry fee paid.
\end{itemize}

\subsection{Feature Representation}
In addition to the rich historical features embedded in the user's interaction sequence $S_u$, our model's ranking decision relies on two other critical categories of features that are specific to each ranking instance.
\subsubsection{\textbf{Temporal Positional Encoding}}
To model the decaying influence of past interactions, we incorporate a temporal positional encoding. 
For each historical action $a_j$ at time $t_j$ from the user's sequence $S_u$, we compute a time-gap feature, $\Delta t_j = t_c - t_j$,
representing the time elapsed between the past action and the current prediction time $t_c$. 
This feature is incorporated into the representation of each historical action before being fed into the model.

\subsubsection{\textbf{Target Item Features}}
Each candidate match $i$ in the slate $\mathcal{C}(u,t_c)$ is represented by a feature vector $\mathbf{z}_i$ that captures its real-time context and urgency. 
This vector includes crucial urgency features, such as Time-To-Round-Lock (TTRL) and Time-Since-Lineups, as well as contextual features like the match's maximum prize amount.

\subsection{Objective Function}
The primary learning task is to approximate a scoring function, $f:(S_u, i, \mathbf{z}_i, t_c) \mapsto \hat{y}_i$, that produces a relevance score $\hat{y}_i$ for each candidate item $i$.
For a given candidate set $\mathcal{C}(u,t_c)$, the final output is a ranked list $\pi$, which is a permutation of $\mathcal{C}(u,t_c)$ created by sorting the items based on their predicted scores [$\hat{y}_1, \hat{y}_2, \ldots, \hat{y}_m$]. 
To train the model to produce a high-quality ranking, our objective is to minimize a listwise loss function that serves as a differentiable surrogate for a desired offline ranking metric, in this case, nDCG. The specifics of our chosen loss function are detailed in the next section.

\section{Proposed Method: An Urgency-Aware Ranking Framework}\label{sec:proposed_methodology}
This section details our proposed deep learning framework for urgency-aware match ranking. 
Our approach adapts the core target-attention principle from the Deep Interest Network (DIN) architecture \cite{din}. 
However, whereas the original DIN was designed for a pointwise Click-Through Rate (CTR) prediction task, we fundamentally re-engineer its architecture for a listwise ranking objective. 
We begin by describing the overall model architecture, followed by specifics of its key components.

\subsection{Model Architecture}
\begin{figure}[t]
    \centering
    \includegraphics[width=0.9\columnwidth]{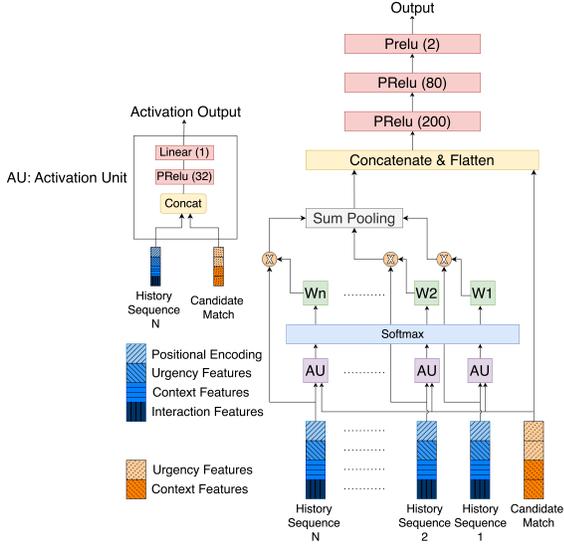}
    \caption{The overall architecture of our proposed Urgency-Aware DIN model. The model computes a target-aware user interest vector via an attention mechanism, which is then combined with item features and fed into a final MLP for ranking.}
    \label{fig:model_architecture}
\end{figure}

Our proposed framework is an adaptation of the Deep Interest Network (DIN) architecture \cite{din}, which we have tailored to our feature-rich, urgency-aware domain. 
The model is designed to compute a relevance score for each candidate match by generating a user interest representation that is dynamically adapted to that specific match. 
The overall architecture is illustrated in Figure \ref{fig:model_architecture}.

The model takes as input the user's historical interaction sequence, $S_u$, and the set of feature vectors $\{\mathbf{z}_1, \ldots, \mathbf{z}_m\}$ for all candidate matches in the slate $\mathcal{C}(u,t_c)$. 
At its core, the model first computes a target-aware user interest vector using an attention mechanism, and then passes this vector along with the target item's features into a final prediction network.

The central innovation of DIN is its \textbf{target-attention mechanism}. 
For each candidate match $i$, this unit computes a tailored user interest vector. 
It achieves this by individually assessing the relevance of each historical interaction $a_j \in S_u$ with respect to the candidate $i$. The feature vector of the historical action and the feature vector of the candidate match, $\mathbf{z}_i$, are concatenated and passed through a small feed-forward network (an activation unit) to produce a scalar attention weight.

The attention weights are then normalized to produce the final attention scores, $\alpha_{ij}$. 
While the original DIN paper proposes a specialized activation unit to scale weights without normalization across the historical items, we found that a standard softmax function provided a robust and effective normalization mechanism in our applied setting. 
To handle variable-length user histories, we apply a mask to the weights of padded items before the softmax operation, ensuring they do not contribute to the final representation. 
The final, target-aware user interest vector, $\mathbf{v}_u(i)$, is then computed as the weighted sum of the historical action vectors, using the attention scores as weights:
\begin{equation}
    \mathbf{v}_u(i) = \sum_{j=1}^{|S_u|} \alpha_{ij} \cdot \mathbf{v}(a_j)
\end{equation}
where $\mathbf{v}(a_j)$ is the input vector representation of the historical action $a_j$. 
Crucially, a distinct user interest vector $\mathbf{v}_u(i)$ is computed for each candidate item $i$.

Finally, this dynamic user interest vector $\mathbf{v}_u(i)$ is concatenated with the target item's own feature vector $\mathbf{z}_i$. 
This combined vector is passed through a deep feed-forward network with several hidden layers and PReLU activations to produce the final scalar relevance score, $\hat{y}_i$.
\begin{equation}
    \hat{y}_i = \text{MLP}(\mathbf{v}_u(i) \oplus \mathbf{z}_i)
\end{equation}
where $\oplus$ denotes vector concatenation.

This entire process is repeated for every candidate item in the slate, producing a list of scores $[\hat{y}_1, \ldots, \hat{y}_m]$ that can be directly used by a listwise loss function for end-to-end training.

\subsection{Listwise Optimization with neuralNDCG}
To align our training objective directly with the business goal of ranking quality, we employ a listwise loss function. Specifically, we use neuralNDCG \cite{neuralndcg}, a state-of-the-art differentiable surrogate for the Normalized Discounted Cumulative Gain (nDCG) metric.

The core challenge in directly optimizing ranking metrics like nDCG is that the sorting operation required to compute rank is non-differentiable. 
neuralNDCG overcomes this by implementing a differentiable relaxation of sorting known as NeuralSort \cite{neural_sort}. 
Conceptually, instead of producing a discrete, hard permutation of the ranked list, NeuralSort uses the model's predicted scores $[\hat{y}_1, \ldots, \hat{y}_m]$ to produce a "soft" permutation matrix, $P_{\hat{y}}$. 
This matrix represents a probability distribution over all possible permutations.

This differentiable permutation matrix allows for the computation of an expected nDCG value for the ranked slate. Our model is trained end-to-end by minimizing the negative of this value, which is equivalent to maximizing the nDCG of the recommended slate:

\begin{equation}
    \mathcal{L}_{\text{rank}} = -\text{neuralNDCG}(\hat{y}, y)
\end{equation}
where $\hat{y}$ is the vector of predicted scores from our DIN model for a given slate, and $y$ is the vector of ground-truth relevance labels for that slate. 
For our implementation, we utilize the deterministic variant of NeuralSort as described in the original work \cite{neural_sort}. 
By using this loss, we directly optimize the parameters of our network to produce higher-quality ranked lists, bridging the gap between the training objective and the final evaluation metric.

\section{Offline Evaluation}\label{sec:offline_evaluation}
\subsection{Dataset}
\begin{table*}[t]
    \centering
    \begin{tabular}{lcccc}
        \toprule
        \textbf{Split} & \textbf{Users} & \textbf{Interactions} & \textbf{Time Period} & \textbf{Purpose} \\
        \midrule
        Training & 200,000 & 92.5B & 12 months & Model Training \\
        Validation & 200,000 & 10.5B & 4 months & Hyperparameter Tuning \\
        Test & 250,000 & 11.7B & 4 months & Final Performance Evaluation \\
        \bottomrule
    \end{tabular}
    \caption{Dataset Statistics}
    \label{tab:dataset_stats}
\end{table*}
Our experiments are conducted on a massive, real-world industrial dataset from the Dream11 Daily Fantasy Sports (DFS) platform. To ensure a rigorous evaluation, we employ a strict disjoint user and out-of-time splitting strategy. We first partition a large cohort of existing users into three non-overlapping sets for training, validation, and testing, with the validation and test periods occurring sequentially after the training period. This setup specifically measures the model's ability to generalize its learned patterns to users whose histories it has not been trained on, a challenging and realistic industrial scenario. The key statistics for each data split are summarized in Table \ref{tab:dataset_stats}.

\subsection{Evaluation Metrics}
We evaluate model performance on the held-out test set using two standard ranking metrics. Recall@k measures the fraction of a user’s positive interactions (match clicks) that appear in the top-k ranked items, and we report it for k={1, 3, 5}. Our primary metric is nDCG@k, which evaluates the quality of the entire ranked list by rewarding higher placement of relevant items. We report nDCG for k={1, 3, 5}. These cutoffs are chosen as they directly reflect the user experience, where a maximum of five match cards are visible on the homepage at any time.

\subsection{Baselines}
We compare our framework against two strong industrial baselines. Our primary baseline is a User-Level LightGBM Ranker, a large-scale model trained on the entire user population using the SynapseML library \cite{synapseML}. As a secondary baseline, we include a Segmented LightGBM Ranker, which trains separate models for two distinct user cohorts to account for behavioral heterogeneity: highly-engaged "Power Users" who exhibit strategic, utility-maximizing patterns, and more casual "Non-Power Users" whose engagement is often driven by major sporting events. Both baselines are gradient boosted decision tree ensembles optimized with the lambdarank objective. They are trained on an extensive set of handcrafted features that mirrors the urgency and contextual features used for the candidate items in our proposed model. The critical difference is that the baselines do not leverage the raw, feature-rich user interaction sequence, which is the primary input for our DIN-based approach. Full hyperparameter details are provided in the appendix.

\subsection{Results and Analysis}
This section presents the results of our offline experiments. 
We first compare the main performance of our proposed model against the baselines, followed by a series of ablation studies to understand the contribution of each model component.

\subsubsection{Main Performance Comparison}
\begin{figure}[t]
    \centering
    \includegraphics[width=0.9\columnwidth]{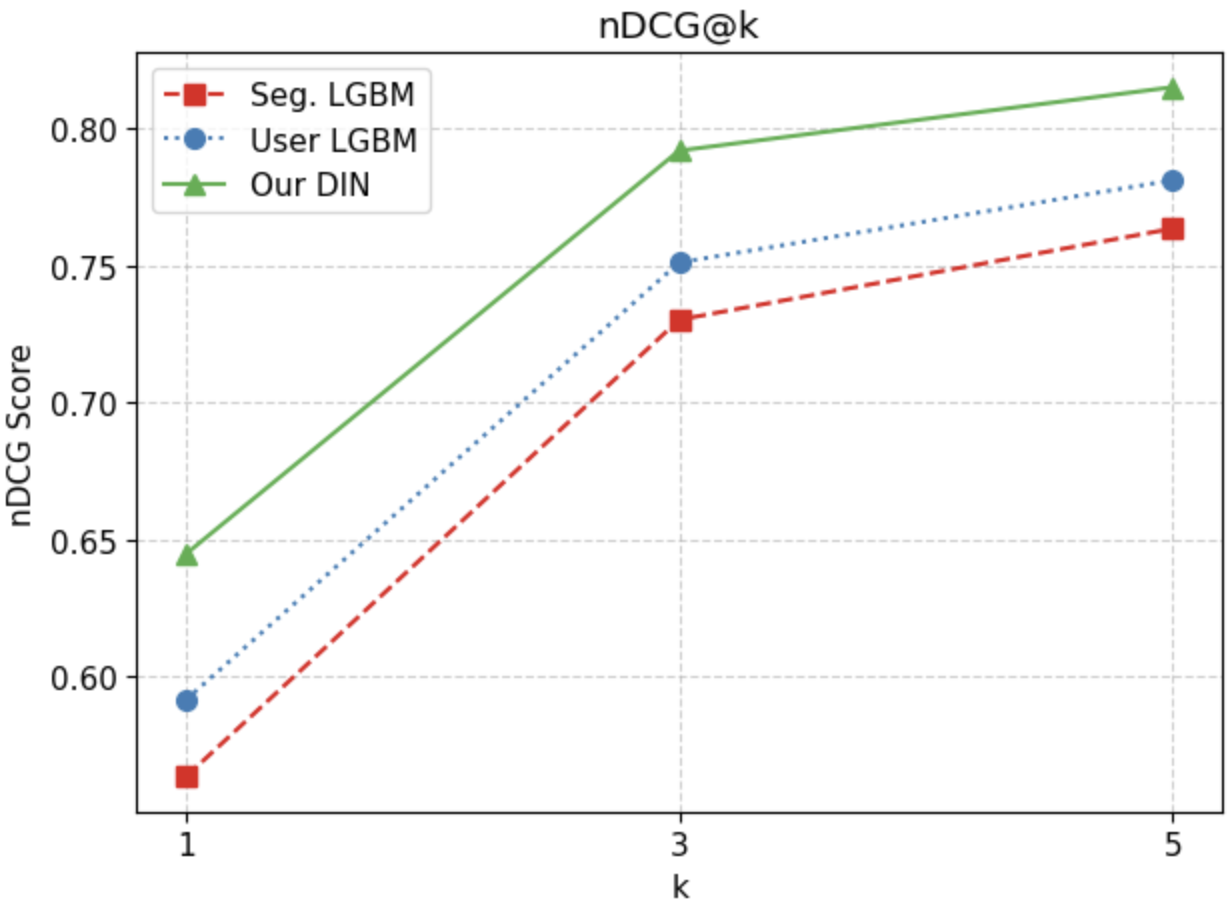}
    \caption{Model Performance Comparison on the held-out test set. Our proposed Urgency-Aware DIN model shows a clear performance lift over both LightGBM baselines across all nDCG@k metrics.}
    \label{fig:performance_chart_nDCG}
\end{figure}
\begin{figure}[t]
    \centering
    \includegraphics[width=0.9\columnwidth]{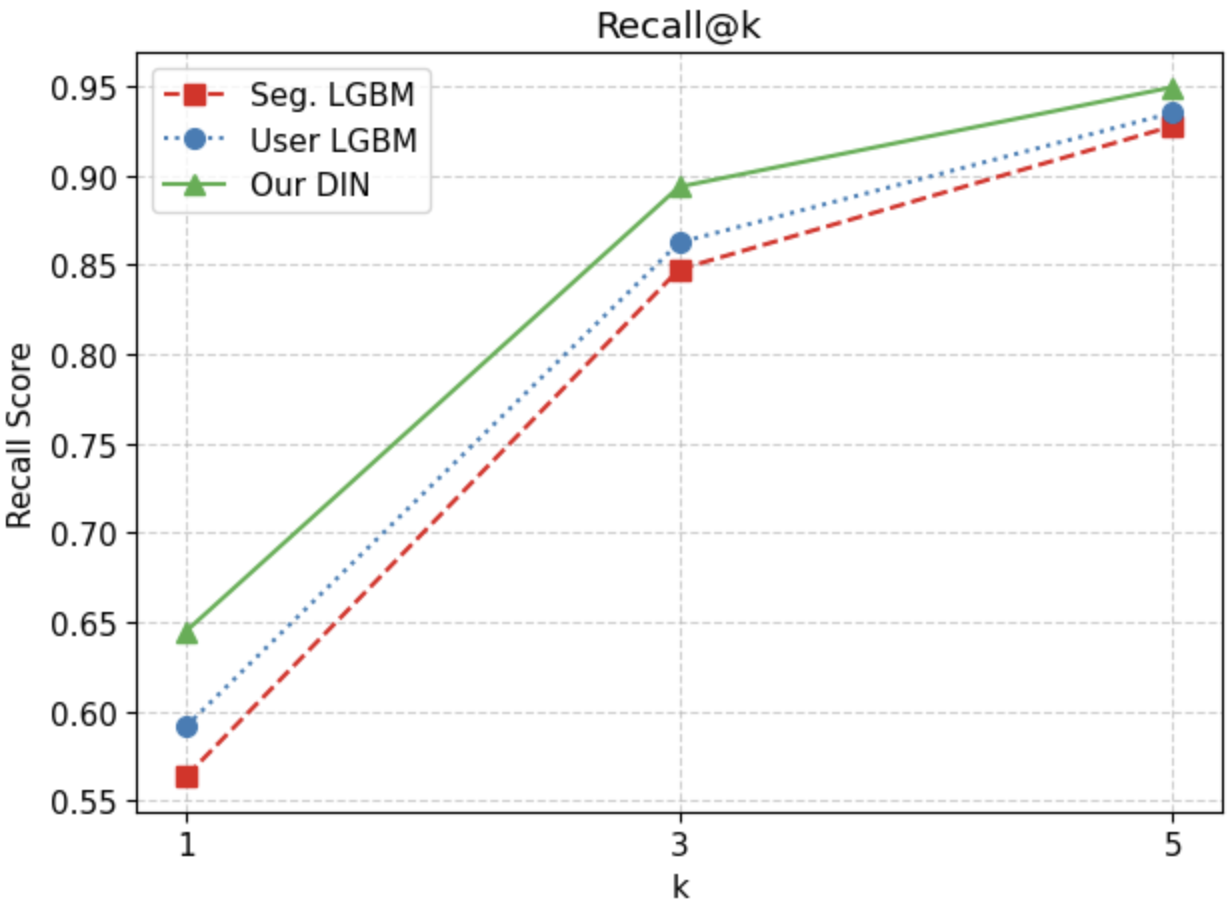}
    \caption{Model Performance Comparison on the held-out test set. Our proposed Urgency-Aware DIN model shows a clear performance lift over both LightGBM baselines across all Recall@k metrics.}
    \label{fig:performance_chart_recall}
\end{figure}
The primary results of our offline evaluation are presented in Figure \ref{fig:performance_chart_nDCG} and Figure \ref{fig:performance_chart_recall}. 
The data clearly demonstrates that our proposed Urgency-Aware DIN model significantly outperforms both the User-Level and Segmented LightGBM baselines across all reported nDCG@k and Recall@k metrics. Notably, our model achieves a +9\% relative lift in nDCG@1 over the primary User-Level LightGBM baseline. This is a particularly impactful result, as improving the top-ranked item is critical for user engagement on the homepage. The performance gains, while decreasing for higher values of k, remain consistent, indicating that our model is not only better at retrieving relevant items but also at placing them in the most prominent positions.
These results validate our central hypothesis: that a deep learning model capable of directly learning from raw user interaction sequences can capture more nuanced user preferences than a GBDT model that relies on an extensive but fixed set of handcrafted features. The DIN architecture's ability to dynamically weigh historical interactions based on the target item allows it to uncover patterns that are difficult to engineer manually.

\subsubsection{Ablation Studies}
\begin{table}[t]
\setlength{\tabcolsep}{1mm}
    \centering
    \begin{tabular}{lccc}
    \toprule
    \textbf{Variant} & \textbf{nDCG@1} & \textbf{nDCG@3} & \textbf{nDCG@5}\\
    \midrule
    \textbf{Full Model} & \textbf{0.6445} & \textbf{0.7920} & \textbf{0.8152}\\
    w/ Pointwise Loss & 0.6405 & 0.7893 & 0.8129\\
    w/o Pos. Encoding & 0.6288 & 0.7812 & 0.8058\\
    w/o Urgency Feats & 0.3832 & 0.5240 & 0.5676\\
    \bottomrule
    \end{tabular}
    \caption{Ablation study results on the test set.}
    \label{tab:ablation_results}
\end{table}
To isolate and understand the contribution of each key component of our framework, we conduct a series of ablation studies. The results, presented in Table \ref{tab:ablation_results}, confirm that each component provides a significant and complementary contribution to the model's overall effectiveness. Most notably, removing the Urgency Features causes the largest performance degradation, validating our central hypothesis that explicitly modeling real-time urgency is critical. The performance drop after removing the Positional Encoding and using a Pointwise Loss further confirms the value of modeling historical recency and employing a listwise optimization objective, respectively.

\section{Large Scale Distributed Training Setup}
\begin{figure*}[t]
    \centering
    \includegraphics[width=1.43\columnwidth]{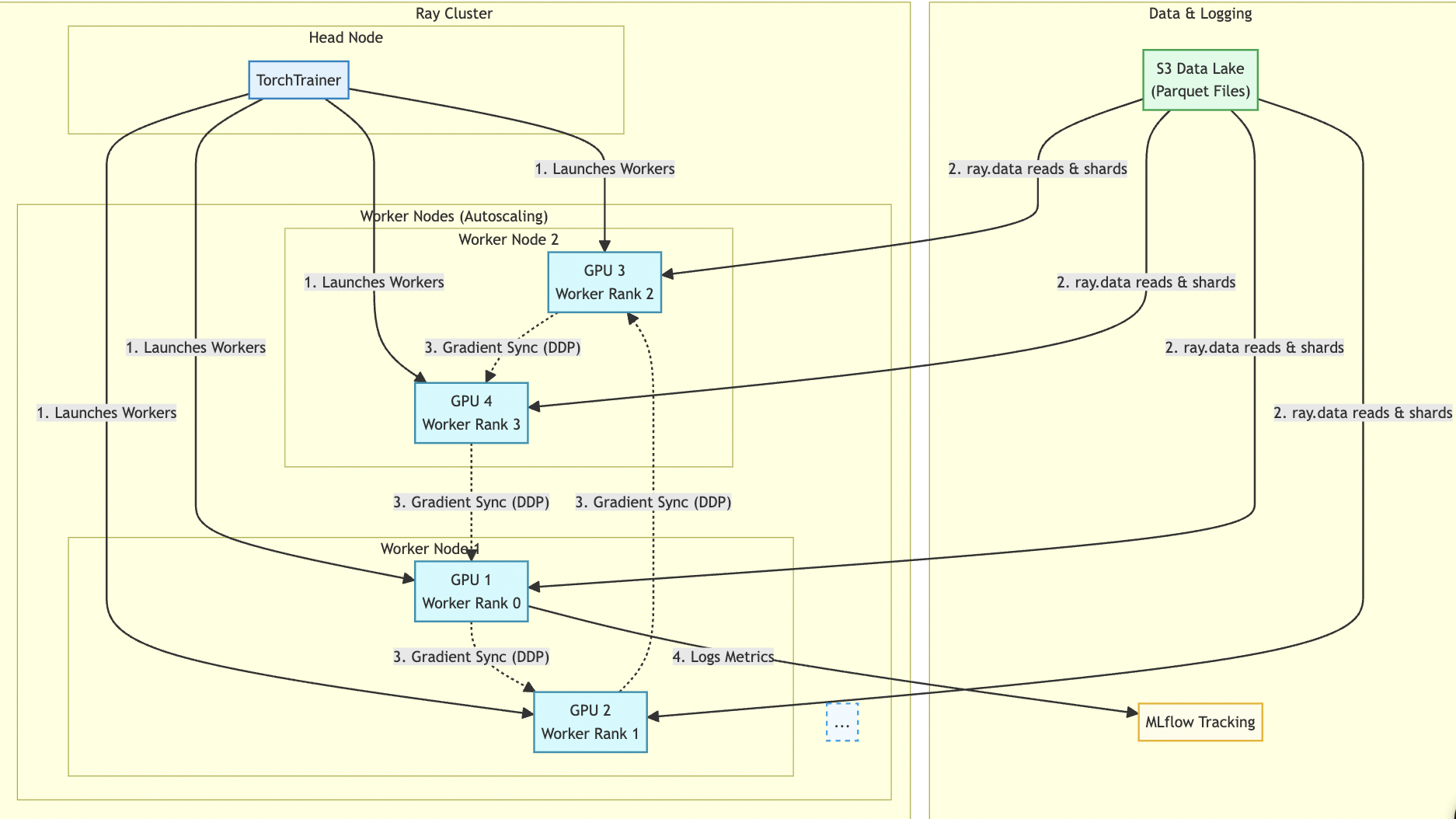}
    \caption{A visual representation of the multi-node, multi-GPU training framework.}
    \label{fig:distributed_training_diagram}
\end{figure*}

To train our model at an industrial scale on over 100 billion interactions, we developed a multi-node, multi-GPU distributed training framework using Ray \cite{ray_paper} and PyTorch \cite{pytorch_paper}, as illustrated in Figure \ref{fig:distributed_training_diagram}. We leverage ray.data to efficiently load and shard our Parquet dataset directly from S3 across the cluster. The training is orchestrated by ray.train.torch.TorchTrainer, which launches a configurable number of workers (e.g., 80 workers for an 80-GPU job) across multiple nodes. Each worker is assigned a dedicated GPU and utilizes PyTorch's Distributed Data Parallel (DDP) for gradient synchronization. This distributed framework was critical for our project, reducing end-to-end training time from an estimated several days to around 1 hour (per epoch), which enabled the rapid experimentation and extensive tuning required for this work.
\section{Path to Deployment: On-Device Ranking}
\begin{figure}[t]
    \centering
    \includegraphics[width=0.9\columnwidth]{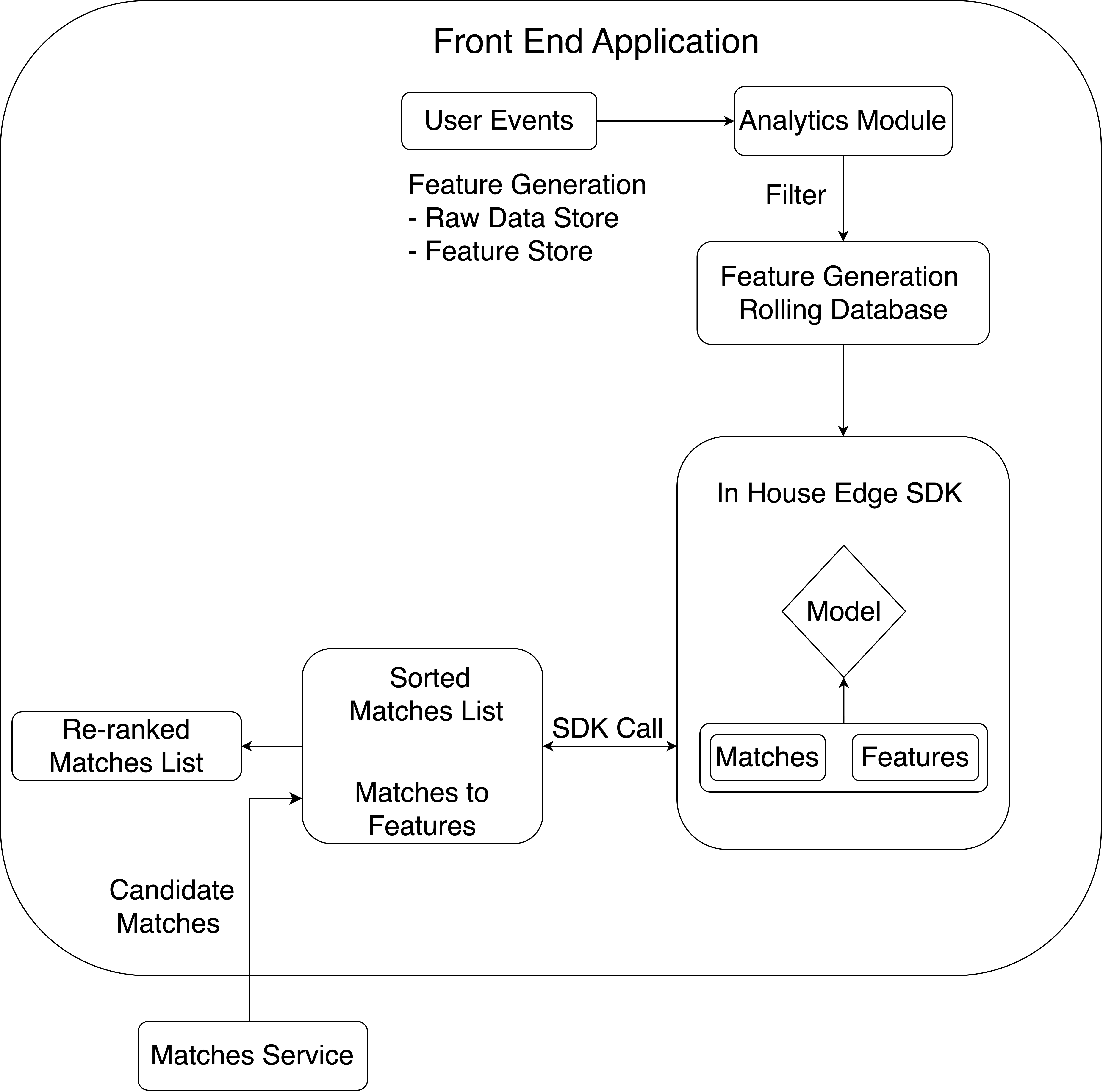}
    \caption{High-level overview of the on-device ranking workflow.}
    \label{fig:edge_architecture}
\end{figure}
This work serves as the foundation for our next-generation, on-device recommendation system (Figure \ref{fig:edge_architecture}). The architecture uses an in-house Edge SDK to enrich a candidate list from a backend service with on-device historical features. Our DIN model then performs low-latency re-ranking locally, which minimizes server load and network dependency. The strong offline results presented here validate this architecture's viability for future online A/B testing.

\appendix
\section{Hyperparameters and Implementation Details}
The hyperparameters for all models were tuned on our held-out validation set. The final parameters used for the test set evaluation are detailed below.
\subsection{LightGBM Baselines}
Our two LightGBM baselines were trained using the lambdarank objective. The key hyperparameters for the User-Level LightGBM model, trained using the SynapseML library, are listed in Table~\ref{tab:lgbm_user_level_hyperparams}. The hyperparameters for the Segmented LightGBM models are listed in Table~\ref{tab:lgbm_segment_hyperparams}.
\subsection{Urgency-Aware DIN Model}
Our proposed Urgency-Aware DIN model was implemented in PyTorch. The key hyperparameters are detailed in Table \ref{tab:din_hyperparams_appendix}.
\subsection{Implementation Details}
Our deep learning models were implemented using PyTorch (v2.6.0). The distributed training framework was built using Ray (v2.37.0). All experiments were conducted on a cloud-based cluster, with each training worker utilizing an NVIDIA A10G GPU. The LightGBM baselines were trained using the SynapseML library on an Apache Spark cluster.
\begin{table}[t]
    \centering
    \begin{tabular}{lcc}
        \toprule
        \textbf{Hyperparameter} & \textbf{Value} \\
        \midrule
        numLeaves & $32$ \\
        numIterations & $500$ \\
        metric & $ndcg$ \\
        objective & $lambdarank$ \\
        \bottomrule
    \end{tabular}
    \caption{Tuned hyperparameters for the User-Level LightGBM model}
    \label{tab:lgbm_user_level_hyperparams}
\end{table}
\begin{table}[t]
    \centering
    \begin{tabular}{lcc}
        \toprule
        \textbf{Hyperparameter} & \textbf{Power Users} & \textbf{Non-Power Users} \\
        \midrule
        objective & $lambdarank$ & $lambdarank$ \\
        metric & $ndcg$ & $ndcg$ \\
        max\_depth & $6$ & $10$ \\
        n\_estimators & $300$ & $200$ \\
        learning\_rate & $0.01$ & $0.05$ \\
        \bottomrule
    \end{tabular}
    \caption{Tuned hyperparameters for the Segmented LightGBM models.}
    \label{tab:lgbm_segment_hyperparams}
\end{table}
\begin{table}[t]
\centering
\begin{tabular}{lc}
\toprule
\textbf{Hyperparameter} & \textbf{Value} \\
\midrule
Optimizer & Adam \\
Learning Rate & 3e-4 \\
Batch Size & 512 \\
Activation Function & PReLU \\
\midrule
\multicolumn{2}{c}{\textbf{Attention Unit}} \\
\midrule
Hidden Layer Size & 32 \\
\midrule
\multicolumn{2}{c}{\textbf{Prediction MLP}} \\
\midrule
Hidden Layer Sizes & [200, 80, 2] \\
\bottomrule
\end{tabular}
\caption{Tuned hyperparameters for our proposed Urgency-Aware DIN model.}
\label{tab:din_hyperparams_appendix}
\end{table}
\bibliography{aaai2026}
\end{document}